\documentstyle[12pt,epsf]{article}
\renewcommand{\bar}[1]{\overline{#1}}

\begin{document}

\begin{flushright}
USM-TH-117
\end{flushright}

\centerline{\Large \bf  $q \to \Lambda$ Fragmentation Function and
Nucleon Transversity}

\centerline{\Large \bf Distribution in a Diquark Model}

\vspace{22pt}
\centerline{\bf
Jian-Jun Yang\footnote{e-mail: jjyang@fis.utfsm.cl}$^{a,b}$}

\vspace{8pt}

{\centerline {$^{a}$Department of Physics, Nanjing Normal
University,}}

{\centerline {Nanjing 210097, China}}

{\centerline {$^{b}$Departamento de F\'\i sica, Universidad
T\'ecnica Federico Santa Mar\'\i a,}}

{\centerline {Casilla 110-V, %\\ \null\quad
Valpara\'\i so, Chile\footnote{Mailing address}}

\vspace{10pt}
\begin{center} {\large \bf Abstract}

\end{center}

Based on a simple quark-diquark model, we propose a set of
unpolarized, longitudinally polarized and transversely polarized
fragmentation functions for the $\Lambda$  by fitting the
unpolarized $\Lambda$ production data in $e^+ e ^- $ annihilation.
It is found that the helicity structure of the obtained $\Lambda$
fragmentation functions is supported by the all available
experimental data on the longitudinal $\Lambda$ polarization.
Within the same framework of the diquark model, the nucleon
transversity distributions are presented and consistent
descriptions of the available HERMES data on the azimuthal spin
asymmetries in pion electroproduction are obtained. Furthermore,
the spin transfers to the transversely polarized $\Lambda$ in the
charged lepton DIS on a transversely polarized nucleon target are
predicted for future experiments.

\vspace{2cm}
\begin{center}
{\bf (Talk given in a seminar of U. Santa Mar\'\i a)}
\end{center}

\vfill

\newpage
\section{Introduction}

In recent years, there has been great progress in understanding
the detailed flavor and spin structure of the nucleon. In order to
enrich our knowledge of the nucleon structure, it is  significant
to apply the same mechanisms that produce the quark structure of
the nucleon to other octet baryons and to find a new domain where
the physics invoked to explain the structure of the nucleon can be
clearly checked.  The $\Lambda$ hyperon is of special interest in
this respect since its polarization can be easily measured in the
usual fashion through the self-analyzing  decay $\Lambda \to p
\pi^-$. Actually, there has been some recent progress in
measurements of polarized $\Lambda$ production. The longitudinal
$\Lambda$ polarization in $e^+e^-$ annihilation at the Z-pole was
observed by several collaborations at CERN [1-3]. Recently, the
HERMES Collaboration at DESY reported a result for the
longitudinal spin transfer to the $\Lambda$ in the polarized
positron deep inelastic scattering (DIS) process~\cite{HERMES}.
Also the E665 Collaboration at FNAL measured the $\Lambda$ and
$\bar{\Lambda}$ spin transfers from muon DIS~\cite{E665}, and they
observed a very different behaviour for $\Lambda$'s and
$\bar{\Lambda}$'s. Very recently, the measurement of $\Lambda$
polarization in charged current interactions has been done in the
NOMAD experiment~\cite{NOMAD}.  On the other hand, much work has
already been done to relate the flavor and spin structure of the
$\Lambda$ to various fragmentation processes [7-25]. One of the
most interesting  observations is related to the polarization of
quarks inside the $\Lambda$. In the naive quark model, the
$\Lambda$ spin is exclusively provided by the strange ($s$) quark,
and the $u$ and $d$ quarks are unpolarized. Based on novel results
concerning the proton spin structure from DIS experiments and
SU(3) symmetry in the baryon octet, it was found that the $u$ and
$d$ quarks of the $\Lambda$ should be negatively
polarized~\cite{Bur93}. However, based on the light cone SU(6)
quark diquark spectator model and
 the perturbative QCD (pQCD) counting rules analysis,
 it was found that  the $u$ and $d$ quarks should be
positively polarized at large $x$, even though their net spin
contributions to the $\Lambda$ might be zero or
negative~\cite{MSY2-3}. Very recent analysis~\cite{YJJ-02} with a
statistical model also supports the above observation. In
consideration of the fact that what one can actually measure in
experiments are the quark to $\Lambda$ fragmentation functions,
some models for the fragmentation functions have been
proposed~\cite{Nza95}. The unpolarized and polarized fragmentation
functions were modeled but lack a close connection to the
experimental data. Along another direction,  the unpolarized
$\Lambda$ fragmentation functions have been
parametrized~\cite{Flo98b} with the assumption of SU(3) flavor
symmetry  and  various sets of polarized $\Lambda$ fragmentation
functions have been proposed~\cite{Flo98b,Kot98,Bor99b} based on a
simple ansatz such as $\Delta D_{q}^{\Lambda}(z) =C_{q}(z)
D_{q}^{\Lambda}(z)$ with some assumed  coefficients $C_{q}(z)$, or
Monte Carlo event generators without a clear physics motivation.
The high statistics investigation of polarized $\Lambda$
production is one of the main future goals of the HERMES
Collaboration which will improve their detector for this purpose
by adding so called Lambda-wheels. It is very timely as more
reliable unpolarized and polarized $\Lambda$ fragmentation
functions with a clear physics motivation are necessary  to serve
as basis for the analysis of the new HERMES data.

Another significant approach of enriching  our knowledge of the
nucleon structure is to extend the analysis from  the unpolarized
and helicity distributions to the transverse distributions of the
nucleon. The unpolarized, helicity and transverse distributions
are three fundamental quark distributions of the nucleon. The
unpolarized and helicity distributions have been known with some
precision both experimentally and theoretically. The quark
transverse distributions of the nucleon are less known since they
are not directly observables  in inclusive DIS processes. Among
the proposals to measure the quark transverse distributions, the
azimuthal asymmetry in semi-inclusive hadron production has been
considered~[26-30] through the Collins effect~\cite{Col93} of
non-zero production between a chiral-odd structure function and a
T-odd fragmentation function. Actually, the HERMES
Collaboration~\cite{HERMES00} has recently reported evidence for
single-spin asymmetries for semi-inclusive pion production in deep
inelastic scattering  of unpolarized positron beam  on the
longitudinally polarized proton target. Significant spin
asymmetries of the distribution in azimuthal angle $\phi$ of the
$\pi^+$ and $\pi^0$ relative to the lepton scattering plane have
been observed. It is found that the azimuthal asymmetries can
provide information on the quark transverse distributions of the
nucleon. In addition,  Artru and Mekhfi~\cite{Art90}, and later
Jaffe~\cite{Jaf96}, have noticed that the $\Lambda$-hyperon
transverse polarization, in the current fragmentation region of
charged lepton DIS on the transversely polarized nucleon target,
can also provide information on the quark transverse distributions
of the target. Thus it is possible to make a systematic study on
the quark transverse distributions of the nucleon and the
polarized $q  \to \Lambda$ fragmentation functions by using the
available facilities, such as COMPASS, HERMES and SMC, for
detecting $\Lambda$ fragmentation in charged lepton DIS on both
longitudinally and transversely polarized nucleon targets. In fact
there is a real need for more realistic predictions for the quark
transverse distributions  of the nucleon for future experiments.

We have noticed that a simple  quark diquark model given in
Ref.~\cite{Nza95} with a clear physics picture can be used to
provide realistic predictions for  the $\Lambda$ fragmentation
functions and the quark transverse distributions of the nucleon
including the SU(3) symmetry breaking effect. We plan to propose a
set of unpolarized, longitudinally polarized and transversely
polarized $q \to \Lambda$  fragmentation functions  based on the
quark-diquark model~\cite{Nza95}. The unpolarized $\Lambda$
fragmentation functions are optimized by a fit to the unpolarized
cross section of the produced $\Lambda$ in $e^+e^-$ annihilation.
Then we relate the polarized $\Lambda$ fragmentation functions to
the unpolarized ones at the initial scale by using some inputs
from the diquark model calculation and check the obtained
fragmentation functions by means of the available measurement
results on the $\Lambda$ polarization. Within the same framework
of the diquark model, the quark transverse distributions of the
nucleon will  also be provided. Furthermore, we predict the
azimuthal spin asymmetry in pion electroproduction and the spin
transfer to the transversely polarized $\Lambda$ in the charged
lepton DIS on a transversely polarized nucleon target.

This paper is a simply extension of our previous work~\cite{a01} 
with a purpose of checking the transversely polarized 
fragmentation functions and nucleon transversity. In Sec.~2, we describe the
quark to $\Lambda$ fragmentation functions based on the
quark-diquark model. In Sec.~3, the comparisons between our
predictions of the longitudinal polarization for the $\Lambda$
produced in various processes and the corresponding available
experimental data  are made. In Sec.~4, the quark transverse
distributions of the nucleon are proposed based on the diquark
model. In Sec.~5, the calculated results of the azimuthal spin
asymmetries  in pion electroproduction are shown to be consistent
with the available HERMES data. In Sec.~6, we predict the spin
transfer to the  transversely  polarized $\Lambda$  in the charged
lepton DIS on a transversely polarized nucleon target. Finally, we
give a brief discussion and summary with our conclusions in
Sec.~7.

\section{$q \to \Lambda$ fragmentation functions}

Theoretically, the quark fragmentation functions appear in the QCD
description of hard processes as the parts that connect the quark
and gluon lines to hadrons in the final state. Following
Ref.~\cite{Nza95},  the fragmentation of a quark into  a specific
hadron ($p$, $\Lambda$, ...) can be modeled with a quark-diquark-hadron
vertex and the form factors for scalar and axial vector
diquark are  taken as the same form

\begin{equation}
\phi (k^2)=N \frac{k^2-m^2_q}{(k^2- \Lambda_0^2)^2}\label{form}
\end{equation}
with a normalization constant $N$ and a mass parameter
$\Lambda_0$. $\Lambda_0=500~\rm{MeV}$ will be adopted in our
numerical calculations. In Eq.~(\ref{form}), $m_q$ and $k$ are
the mass and  the momentum of the fragmenting quark $q$, respectively.

Within the framework of the diquark  model~\cite{Nza95}, the
unpolarized valence quark to $\Lambda$ fragmentation functions can
be expressed as

\begin{equation}
D_{u_v}^\Lambda(z)=D_{d_v}^\Lambda(z)= \frac{1}{12} a_S^{(u)}(z)
+ \frac{1}{4} a_V^{(u)}(z),
\end{equation}

\begin{equation}
D_{s_v}^\Lambda(z)= \frac{1}{3} a_S^{(s)}(z),
\end{equation}
where $a_D^{(q)}(z)$  ($D=S$ or $V$) is the probability of finding
a quark $q$ splitting into $\Lambda$ with longitudinal momentum
fraction $z$ and emitting a scalar ($S$) or axial vector ($V$)
antidiquark. The SU(3) symmetry breaking effect can be reflected by
considering the mass difference between the scalar diquark and
vector diquark states.

According to Ref.~\cite{Nza95},  $a_D^{(q)}(z)$ can be
expressed in the quark-diquark model as

\begin{equation}
a_D^{(q)}(z)=\frac{N^2 z^2 (1-z)^3}{64 \pi^2}
\frac {[2 ( M_\Lambda + m_q z)^2+
 R^2(z)]}{R^6(z)}
\end{equation}
with

\begin{equation}
R(z)=\sqrt{z m_D^2-z(1-z)\Lambda_0^2+(1-z)M_\Lambda^2},
\end{equation}
where $M_\Lambda$ and $m_D$~($D=S$ or $V$) are the mass of
the $\Lambda$ and a diquark,
respectively. In consideration of the mass difference
$M_\Lambda- M_p= 176~ \rm{MeV}$,
we choose the diquark mass $m_S=900~ \rm{MeV}$ and $m_V=1100~ \rm{MeV}$
for non-strange diquark states, $m_S=(900+176)~ \rm{MeV}$ and
$m_V=(1100+176)~ \rm{MeV}$ for diquark states  $(qs)$  with $q=u,~ d$.
The quark masses are taken as $m_u=m_d= 350~ \rm{MeV}$ and $m_s=(350+176)~
\rm{MeV}$.

Similarly, the longitudinally and transversely polarized
quark to $\Lambda$ fragmentation functions can
be written  as

\begin{equation}
\Delta D_{u_v}^\Lambda(z)=\Delta D_{d_v}^\Lambda(z)
= \frac{1}{12} \tilde{a}_S^{(u)}(z)
- \frac{1}{12} \tilde{a}_V^{(u)}(z),
\end{equation}

\begin{equation}
\Delta D_{s_v}^\Lambda(z)= \frac{1}{3} \tilde{a}_S^{(s)}(z),
\label{Dsv}
\end{equation}
and
\begin{equation}
\delta D_{u_v}^\Lambda(z)=\delta D_{d_v}^\Lambda(z)
= \frac{1}{12} \hat{a}_S^{(u)}(z)
- \frac{1}{12} \hat{a}_V^{(u)}(z),
\end{equation}

\begin{equation}
\delta D_{s_v}^\Lambda(z)= \frac{1}{3} \hat{a}_S^{(s)}(z),
\label{Dsvt}
\end{equation}
respectively, with

\begin{equation}
\tilde{a}_D^{(q)}(z)=\frac{N^2 z^2 (1-z)^3}{64 \pi^2}
\frac {[2 ( M_\Lambda + m_q z)^2-
 R^2(z)]}{R^6(z)},
\end{equation}
and
\begin{equation}
\hat{a}_D^{(q)}(z)=\frac{N^2 z^2 (1-z)^3}{32 \pi^2}
\frac { ( M_\Lambda + m_q z)^2}{R^6(z)}
\end{equation}
for $D=S$ or $V$. What we are interested is not the
magnitude values of the fragmentation functions but the flavor
and spin structure of them which are given by the diquark model.
In order to extract information on the flavor and spin structure,
we introduce the flavor structure ratios

\begin{equation}
F_S^{(u/s)}(z)=\frac{a_S^{(u)}(z)}{a_S^{(s)}(z)},
\end{equation}

\begin{equation}
F_M^{(u/s)}(z)=\frac{a_V^{(u)}(z)}{a_S^{(s)}(z)},
\end{equation}
and the spin structure ratios

\begin{equation}
\tilde{W}_D^{(q)}(z)=\frac{\tilde{a}_D^{(q)}(z)}{a_D^{(q)}(z)},
\end{equation}

\begin{equation}
\hat{W}_D^{(q)}(z)=\frac{\hat{a}_D^{(q)}(z)}{a_D^{(q)}(z)},
\end{equation}
with $D=S$ or  $V$.
Then we can use the unpolarized fragmentation
function $D_{s_v}^\Lambda (z)$ to express all
other unpolarized and polarized fragmentation functions as follows

\begin{equation}
D_{u_v}^\Lambda (z)=[\frac34 F_M^{(u/s)}(z) + \frac14 F_S^{(u/s)}(z) ]
D_{s_v}^\Lambda (z),
\end{equation}

\begin{equation}
\Delta D_{u_v}^\Lambda (z)=\frac 14 [ \tilde{W}_S^{(u)}(z)
F_S^{(u/s)}(z) - \tilde{W}_V^{(u)}(z)  F_M^{(u/s)}(z) ]
D_{s_v}^\Lambda (z),
\end{equation}

\begin{equation}
\Delta D_{s_v}^\Lambda (z)= \tilde{W}_S^{(s)}(z) D_{s_v}^\Lambda
(z),
\end{equation}

\begin{equation}
\delta D_{u_v}^\Lambda (z)=\frac 14 [ \hat{W}_S^{(u)}(z)  F_S^{(u/s)}(z) -
\hat{W}_V^{(u)}(z)  F_M^{(u/s)}(z) ]
D_{s_v}^\Lambda (z),
\end{equation}
and

\begin{equation}
\delta D_{s_v}^\Lambda (z)= \hat{W}_S^{(s)}(z) D_{s_v}^\Lambda (z).
\end{equation}
In this way, the polarized quark to $\Lambda$ fragmentation
functions are related to the unpolarized fragmentation functions
by means of some input ratios from the diquark model calculation.

The quark-diquark description of the fragmentation function should
be reasonable in large $z$ region where the valence quark
contributions dominate. In small $z$ region, the sea contribution
is difficult to be included in the framework of the diquark model
itself. In order to optimize the shape of fragmentation functions,
we adopt simple functional forms

\begin{equation}
D_{s_v}^\Lambda (z) = N_s z ^{\alpha_s} (1-z) ^{\beta_s} \label{fit1}
\end{equation}
and

\begin{equation}
D_{q_s}^\Lambda (z) = D_{\bar{q}}^\Lambda (z)= \bar{N} z ^{\bar{\alpha}}
(1-z) ^{\bar{\beta}} \label{fit2}
\end{equation}
to parametrize fragmentation functions of the valence quark
$D_{s_v}^\Lambda$, sea quark $D_{q_s}^\Lambda (z)$ and antiquark
$D_{\bar{q}}^\Lambda (z)$ for $q=u, d, s$. We assume that
 $D_g^\Lambda$, $\Delta D_g^\Lambda$, $\Delta D_{q_s}^\Lambda$,
 $\Delta D_{\bar{q}}^\Lambda$, $\delta D_{q_s}^\Lambda$, and
 $\delta D_{\bar{q}}^\Lambda$ at the initial scale are zero and they may appear
 due to the QCD evolution. Hence, the input unpolarized and polarized
 $q \to \Lambda$ fragmentation functions  can be written as

\begin{equation}
D_{q}^\Lambda (z) = D_{q_v}^\Lambda (z) + D_{q_s}^\Lambda (z),
\end{equation}

\begin{equation}
\Delta D_{q}^\Lambda (z) = \Delta D_{q_v}^\Lambda (z),
\end{equation}
and

\begin{equation}
\delta D_{q}^\Lambda (z) = \delta D_{q_v}^\Lambda (z).
\end{equation}

\begin{figure}[htb]
\begin{center}
\leavevmode {\epsfysize=10.5cm \epsffile{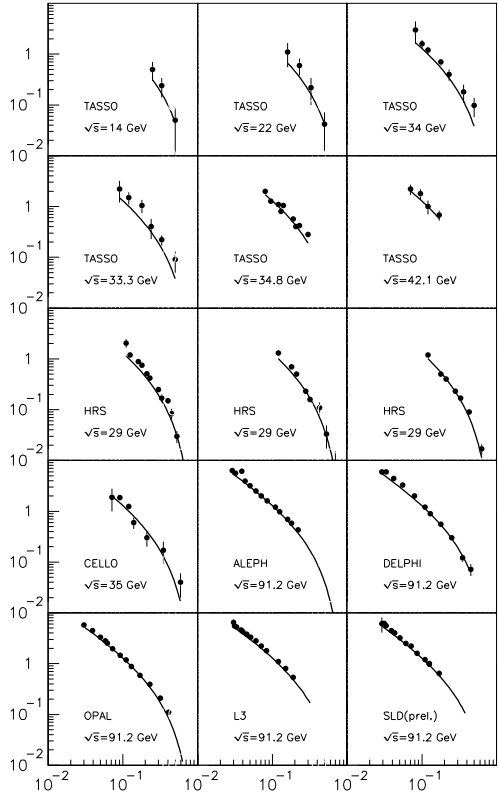}}
\end{center}
\caption[*]{\baselineskip 13pt The comparison of our results for
the $x_E$ dependence of the inclusive $\Lambda$ production cross
section $(1/\sigma_{tot})d \sigma/d x_E$ in $e^+e^-$ annihilation
and the experimental data~[35-40].
%\cite{DELPHI93, ALEPH, TASSO95, OPAL97b, L397, SLD99}
} \label{a03f1}
\end{figure}

\begin{figure}[htb]
\begin{center}
\leavevmode {\epsfysize=4.5cm \epsffile{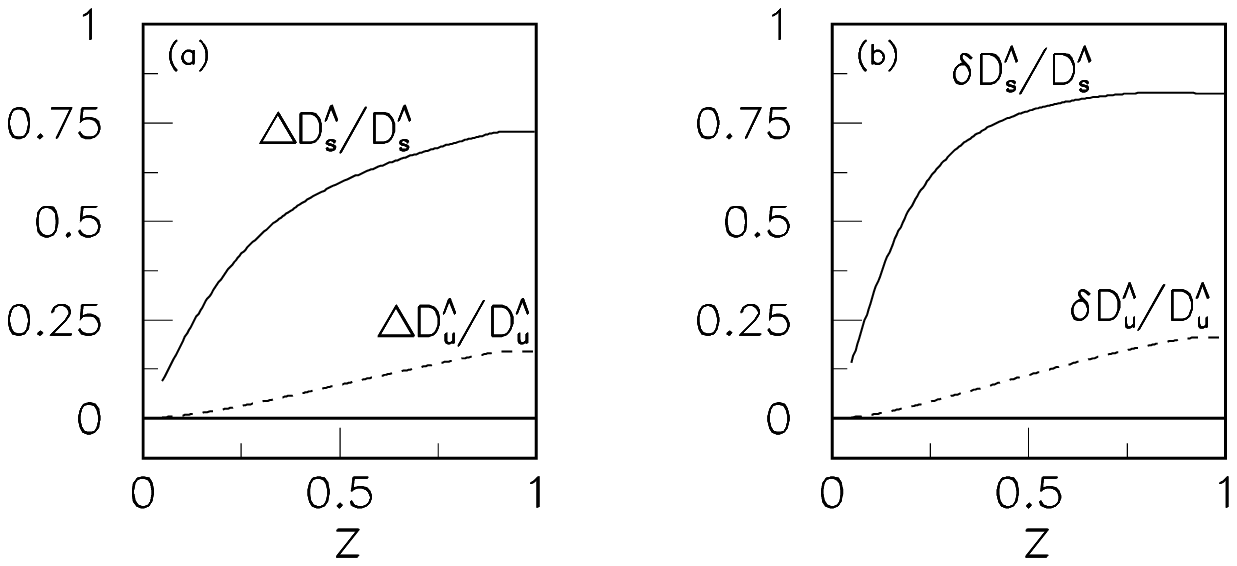}}
\end{center}
\caption[*]{\baselineskip 13pt The spin structure of the $\Lambda$
fragmentation functions at $Q^2=4~\rm{GeV}^2$: (a)~$\Delta
D_q^\Lambda/D_q^\Lambda$ and  (b)~$\delta
D_q^\Lambda/D_q^\Lambda$.} \label{a03f2}
\end{figure}

For a fit to the experimental data, the fragmentation functions
have to be evolved to the scale of the experiments. We use the
evolution package of Ref.~\cite{Miyama94} suitable modified for
the evolution of fragmentation functions in leading order, taking
the input scale $Q_0^2=M_\Lambda^2$ and the QCD scale parameter
$\Lambda_{QCD}= 0.3~ \rm{GeV}$. Furthermore, in order to express
the inclusive cross section and polarization for the  $\Lambda$
production in $e^+e^-$ annihilation , we introduce the following
quantities

\begin{equation}
\hat{A}_q=2 \chi_{2}(v_e^2+a_e^2)v_qa_q-2 e_q \chi_1 a_q
v_e,\label{hatA}
\end{equation}

\begin{equation}
\hat{C}_q=e_q^2-2 \chi_1 v_e v_q e_q+ \chi_2 (a_e^2+v_e^2)
(a_q^2+v_q^2),\label{hatC}
\end{equation}
with
\begin{equation}
\chi_1=\frac{1}{16 \sin^2 \theta_W \cos^2 \theta_W}
\frac{s(s-M_Z^2)}{(s-M_Z^2)^2+M_Z^2\Gamma_Z^2},
\end{equation}

\begin{equation}
\chi_2=\frac{1}{256 \sin^4 \theta_W \cos^4 \theta_W}
\frac{s^2}{(s-M_Z^2)^2+M_Z^2\Gamma_Z^2},
\end{equation}
\begin{equation}
a_e=-1
\end{equation}
\begin{equation}
v_e=-1+4 \sin^2 \theta_W
\end{equation}
\begin{equation}
a_q=2 T_{3q},
\end{equation}
\begin{equation}
v_q=2 T_{3q}-4 e_q \sin^2 \theta_W,
\end{equation}
where $T_{3q}=1/2$ for $u$, while $T_{3q}=-1/2$ for $d$, $s$
quarks, $N_c=3$ is the color number, $e_q$ is the charge of the
quark in units of the proton charge, $s$ is the total
center-of-mass (c.m.) energy squared, $\theta$ is the angle
between the outgoing quark and the
incoming electron, $\theta_W$ is the Weinberg angle, and $M_Z$ and
$\Gamma_Z$ are the mass and width of $Z^0$. In the quark-parton
model, the differential cross section for the semi-inclusive
$\Lambda$ production process $e^+e^- \to \Lambda + X$ can be
expressed  to leading order
\begin{equation}
\frac{1}{\sigma_{tot}}\frac{d \sigma}{d x_E} =\frac{ \sum\limits_q
\hat{C}_q \left [ D_q^\Lambda (x_E,Q^2)+D_{\bar{q}}^\Lambda
(x_E,Q^2) \right ]} {\sum\limits_q \hat{C}_q} \label{crosection}
\end{equation}
where  $x_E=2 E_\Lambda/\sqrt{s}$ with  $E_\Lambda$ being the
energy of the produced $\Lambda$ in the $e^+e^-$ c.m. frame, and
$\sigma_{tot}$ is the total cross section for the process.

We perform  the leading order fit since the analysis in
Ref.~\cite{Flo98b} shows that the leading order fit can arrive at
the same fitting quality as  the next-to-leading order fit.
 By fitting the inclusive  unpolarized $\Lambda$ production data in $e^+e^-$
annihilation, the optimal  parameters in
Eqs.~(\ref{fit1})-(\ref{fit2}) are obtained as  $N_s=2.5$,
$\alpha_s=0.7$, $\beta_s=3.0$, $\bar{N}=0.4$,
$\bar{\alpha}=-0.51$, and $\bar{\beta}=7.4$. In Fig.~\ref{a03f1},
we show our fit results compared with the experimental
data~[35-40]. In addition, we show the spin structure of the
$\Lambda$ fragmentation functions in Fig.~\ref{a03f2}. From
Fig.~\ref{a03f2}, one can see that the transverse and helicity
structure of the $\Lambda$ fragmentation functions in the diquark
model are very similar each other.

\section{Longitudinal $q \to \Lambda$ polarization transfers}

In order to check the obtained $\Lambda$ fragmentation functions,
we apply them to predict the  spin observables in
various $\Lambda$ production processes.

\subsection{$\Lambda$ polarization in $e^+e^-$ annihilation}

Due to the interference between the vector and axial vector
coupling in the standard model for electroweak interaction, quarks
produced in $e^+e^-$-annihilation at the Z-pole are polarized. The
polarization of the initial quarks before the hadronization can be
given according to the standard electroweak theory. With the $q
\to \Lambda$ fragmentation functions,  the $\Lambda$-polarization
in $e^+e^-$-annihilation can be expressed as

\begin{equation}
P^{\Lambda}=-\frac{\sum\limits_{q} \hat{A}_q [\Delta D_q^\Lambda
(z)-\Delta D_{\bar q}^\Lambda (z)]}{\sum\limits_{q} \hat{C}_q
[D_q^\Lambda (z)+ D_{\bar q}^\Lambda (z)]}. \label{PL2}
\end{equation}
where $\hat{A}_q$ and $\hat{C}_q$ ($q=u, d$ and $s$) are given in
(\ref{hatA}) and (\ref{hatC}), respectively. At the Z-pole, 
we have the values for $\hat{A}_u \simeq 128$,
$\hat{A}_d=\hat{A}_s \simeq 231$, $\hat{C}_u \simeq 192$, and
$\hat{C}_d=\hat{C}_s \simeq 247$. In addition, one can  see from
Fig.~\ref{a03f2} that $\Delta D_s^\Lambda (z) /D_s^\Lambda (z)$ is
much larger than $\Delta D_u^\Lambda (z) /D_u^\Lambda (z)=\Delta
D_d^\Lambda (z) /D_d^\Lambda (z)$. Therefore, the polarization of
the produced $\Lambda$ at the Z-pole is mainly controlled  by the
polarized $s\to \Lambda$ fragmentation, i.e., $C_s(z)=\Delta
D_s^\Lambda (z) /D_s^\Lambda (z)$. In the present analysis,
$C_s(z)$ is proposed based on the diquark model with a clear
physics motivation. Therefore, one can contrast the qualitative
features of the spin observable in this process with the
fragmentation functions as shown in Fig.~\ref{a03f2}. Our
theoretical prediction for the $\Lambda$ polarization at the
$Z$-pole is shown in Fig.~\ref{a03f3}(a) together with the
experimental data. From Fig.~\ref{a03f3}(a), we can see that the
calculated $\Lambda$ polarization is consistent with the
experimental data, which indicates that the helicity structure of
the $\Lambda$ fragmentation functions proposed in the diquark
model with a clear physics motivation seems to be more realistic
than that with some limit assumptions~\cite{Flo98b}.

\begin{figure}[htb]
\begin{center}
\leavevmode {\epsfysize=4.5cm \epsffile{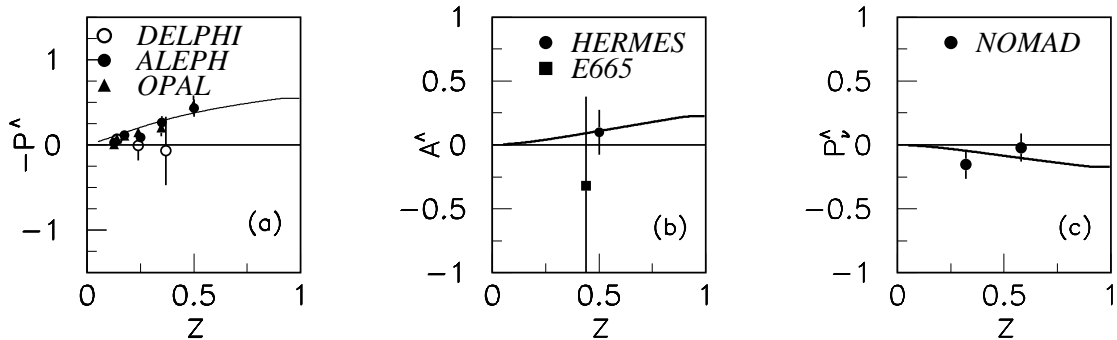}}
\end{center}
\caption[*]{\baselineskip 13pt (a) The comparison of the experimental
data~[1-3]
%\cite{ALEPH96,DELPHI95,OPAL97}
for the longitudinal $\Lambda$-polarization $P^{\Lambda}$ in
$e^+e^-$-annihilation  at the $Z$-pole with our theoretical
prediction; (b) The $z$-dependence of the $\Lambda$ spin transfer
in electron or positron (muon) DIS; (c) The  $z$-dependence of the
$\Lambda$  polarization in the neutrino DIS process. }
\label{a03f3}
\end{figure}

\subsection{Spin transfer to $\Lambda$  in  polarized charged
lepton DIS}

In the charged lepton deeply inelastic scattering process, if the
initial charged lepton beam is polarized, the outgoing quark
should also  be polarized. The polarization of the outgoing quark
depends on $P_B$, the polarization of the initial lepton beam.
There should be a factor $D(y)$ which relates the polarization of
the outgoing quark with that of the initial lepton. The factor
$D(y)$ with its explicit expression
\begin{equation}
D(y)=\frac{1-(1-y)^2}{1+(1-y)^2},
\end{equation}
is commonly referred to as the longitudinal depolarization factor
of the virtual photon with respect to the parent lepton. Here
$y=\nu/E$ is the fraction of the incident lepton's energy that is
transferred to the hadronic system by the virtual photon.
Therefore, for a longitudinally polarized charged lepton beam and
an unpolarized target, the $\Lambda$ polarization along its own
momentum axis is given in the quark parton model by~\cite{Jaf96}
\begin{equation}
P_{\Lambda}(x,y,z) = P_B D(y)A^{\Lambda}(x,z)~,
\label{PL}
\end{equation}
where

\begin{equation}
A^{\Lambda}(x,z)= \frac{\sum\limits_{q} e_q^2 [q^N(x,Q^2) \Delta
D_q^\Lambda(z,Q^2) + ( q \rightarrow \bar q)]} {\sum\limits_{q}
e_q^2 [q^N (x,Q^2) D^\Lambda_q(z,Q^2) + ( q \rightarrow \bar q)]}~
\label{DL}
\end{equation}
is the longitudinal spin transfer to the $\Lambda$. In
Eq.~(\ref{DL}), $q^N(x,Q^2)$,  the quark distribution of the
proton, is adopted as  the CTEQ5 set 1 parametrization
form~\cite{CTEQ5}
 in our numerical calculations.
Our prediction result is shown in  Fig.~\ref{a03f3}(b).
 Note that for HERMES data~\cite{HERMES} the $\Lambda$ polarization is measured along the
virtual-photon momentum, whereas for E665 data~\cite{E665} it is
measured along the virtual-photon spin. The averaged value of the
Bjorken variable is chosen as $x=0.1$ (corresponding to the HERMES
averaged value) and the calculated result is not sensitive to a
different choice of $x$ in the small $x$ region (for example,
$x=0.005$ corresponding to the E665 averaged value).
$Q^2=4~\rm{GeV}^2$ is used and the $Q^2$ dependence of the result
is very weak. The prediction is compatible with the available
experimental data in the medium $z$ region, which suggests that
the $u$ and $d$ quark to the $\Lambda$ fragmentation functions are
likely positive polarized in the medium and large $z$ region. This
prediction is consistent with our previous results based on the
light cone SU(6) quark diquark spectator model and  the pQCD
counting rules analysis~\cite{MSY2-3}, and the statistical
model~\cite{YJJ-02}. However, the present prediction shows the
weaker $z$ dependence in the large $z$ region than the results in
Ref.~\cite{MSY2-3}. We will show in the next subsection that the
NOMAD data~\cite{NOMAD} on the $\Lambda$ polarization in neutrino
DIS seems to favor the weaker $z$ dependence prediction.

Actually, the Lambdas seen at HERMES arise not only from the
fragmentation of $u$, $d$, or $s$ quarks, but also from
electroproduction of heavier hyperons which subsequently decay to
Lambdas. The main contributions from decays are the $\Sigma^0 \to
\Lambda \gamma$ and $\Sigma^*\to \Lambda \pi$. The $\Sigma^0$
decays electromagnetically. Its decay can not be included in any
strong interaction fragmentation function and the $\Lambda$'s
produced via its production and decay must be considered
separately.
 It has been shown that the produced $\Lambda$ which  comes from
 the decay of a heavier hyperon  is also polarized if
 the heavier hyperon  was polarized before its decay. In our present
 analysis, we only included
 the directly produced $\Lambda$.  The contributions from the
 heavier hyperon decay  are
not easily included theoretically at the moment although they
would change the interpretation of the observed polarization in
this process. However, we have  noticed that the electroproduction
of  hyperons  is dominated from the $u$ quark due to the charge
factor for the $u$ quark and that $u^\uparrow (ds)_{0,0}\to
\Lambda$, $u^\uparrow (ds)_{0,0}\to \Sigma^0$ and $u^\uparrow
(ds)_{1,0}\to\Sigma^*$ predicted by the  diquark model have the
similar $z$ dependence.
 So it is expected that the spin transfer in charged lepton DIS
 with the effect of the $\Sigma^0$ and $\Sigma^*$ decays  can be equivalently 
 expressed by the direct hadronized $\Lambda$ spin transfer multiplied by a
factor. Hence, the qualitative feature of the predicted spin
transfer is also expected  to be retained after the heavier
hyperon decay contributions  are further included.

\subsection{$\Lambda$ polarization in neutrino DIS}

One advantage of the neutrino (anti-neutrino) DIS process is that
the scattering of a neutrino beam on a hadronic target provides a
source of polarized quarks with specific flavor structure.
According to the weak interaction theory, the charged current weak
interaction selects only left-handed quarks (right-handed
antiquarks), so the polarization of the fragmenting quarks is
$P_q=-1$. The above  particular property makes the neutrino
process an ideal laboratory to study the flavor-dependence of
quark to hadron fragmentation functions, especially in the
polarized case. We find that the $\Lambda$ polarization in the
neutrino DIS process can also be used to extract information on
the  $u\to \Lambda$ polarization transfer.

The longitudinal polarizations of the $\Lambda$  in its momentum
direction, for the  $\Lambda$ in the current fragmentation region,
can be expressed as~\cite{MSSY5}

\begin{equation}
P_\nu^\Lambda (x,y,z)=-\frac{[d^N(x)+\varpi s^N(x)] \Delta D
_u^\Lambda (z) -( 1-y) ^2 \bar{u}^N (x) [\Delta D
_{\bar{d}}^\Lambda (z)+\varpi \Delta D_{\bar{s}}^\Lambda (z)]}
{[d^N(x)+\varpi s^N(x)] D_u ^\Lambda (z) + (1-y)^2 \bar{u}^N (x)
[D _{\bar{d}}^\Lambda (z)+\varpi D_{\bar{s}}^\Lambda
(z)]}~,\label{neu1}
\end{equation}
where the terms with the factor $\varpi=\sin^2 \theta_c/\cos^2
\theta_c$ ($\theta_c$ is the Cabibbo angle) represent Cabibbo
suppressed contributions. The NOMAD data~\cite{NOMAD} on the
$\Lambda$ polarization in the neutrino DIS process, which has much
better precision than the data on the  longitudinal spin transfer
to the $\Lambda$  in polarized charged lepton DIS, puts further
constraints on the allowed form of the fragmentation functions,
primarily for $u \to \Lambda$ transitions.

In Fig.~\ref{a03f3}(c), we present our prediction of the
$z$-dependence of the $\Lambda$ polarization in the neutrino DIS
process. In the obtained result, we adopted the CTEQ5 set 1 quark
distributions~\cite{CTEQ5} for the target proton at
$Q^2=4$~GeV$^2$ with the Bjorken variable $x$ integrated over
$0.02 \to 0.4$ and $y$ integrated over $0 \to 1$.  The very recent
NOMAD data~\cite{NOMAD} seems to support our prediction that the
$z$-dependence of $\Lambda$ polarization in neutrino DIS is not
strong.

\section{Nucleon transversity distributions}

Within the framework of the quark diquark model~\cite{Nza95}, if
any one of the quarks is probed, re-organize the other two quarks
in terms of two quark wave functions with spin 0 or 1 (scalar and
vector diquarks), {\it i.e.}, the diquark serves as an effective
particle which is called the spectator. Some non-perturbative
effects such as gluon exchanges  in the hadronic debris can be
effectively taken into account by the mass of the diquark
spectator.  More explicitly, the unpolarized valence quark
distributions in the proton can be expressed as

\begin{equation}
u_v^N(x)= \frac{1}{2} A_S^{(u)}(x) + \frac{1}{6} A_V^{(u)}(x),
\end{equation}

\begin{equation}
d_v^N(x)= \frac{1}{3} A_V^{(d)}(x),
\end{equation}
where $A_D^{(q)}(x)$  ($D=S$ or $V$), which is the probability of finding
a quark $q$ to be scattered while the spectator is in the diquark state $D$,
can be expressed in the quark-diquark model as~\cite{Nza95}

\begin{equation}
A_D^{(q)}(x)=\frac{N^2 (1-x)^3}{32 \pi^2}
\frac {[2 ( x M_p + m_q )^2+
 \hat{R}^2(x)]}{\hat{R}^6(x)}
\end{equation}
with

\begin{equation}
\hat{R}(x)=\sqrt{\Lambda_0^2(1-x)+x m_D^2-x(1-x) M_p^2},
\end{equation}
where $M_p$ and $m_D$~($D=S$ or $V$) are the mass of the proton
and a diquark, respectively. Similarly, the polarized quark
distributions of the proton can be written  as

\begin{equation}
\Delta u_v^N(x)= \frac{1}{2} \tilde{A}_S^{(u)}(x) - \frac{1}{18}
\tilde{A}_V^{(u)}(x),
\end{equation}

\begin{equation}
\Delta d_v^N(x)= - \frac{1}{9} \tilde{A}_V^{(d)}(x),
\end{equation}
for the helicity distributions, and

\begin{equation}
\delta u_v^N(x)= \frac{1}{2} \hat{A}_S^{(u)}(x) - \frac{1}{18}
\hat{A}_V^{(u)}(x),
\end{equation}

\begin{equation}
\delta d_v^N(x)= - \frac{1}{9} \hat{A}_V^{(d)}(x),
\end{equation}
for the transverse distributions with

\begin{equation}
\tilde{A}_D^{(q)}(x)=\frac{N^2 (1-x)^3}{32 \pi^2} \frac {[2 ( x
M_p + m_q )^2-
 \hat{R}^2(x)]}{\hat{R}^6(x)}
\end{equation}
and

\begin{equation}
\hat{A}_D^{(q)}(x)=\frac{N^2 (1-x)^3}{16 \pi^2} \frac {( x M_p +
m_q )^2}{\hat{R}^6(x)}
\end{equation}
for $D=S$ or $V$. A detailed analysis of the helicity
distributions has been done in Ref.~\cite{Nza95}. Here, we focus
our attention on the quark transverse distributions of the
nucleon.

\begin{figure}[htb]
\begin{center}
\leavevmode {\epsfysize=4.5cm \epsffile{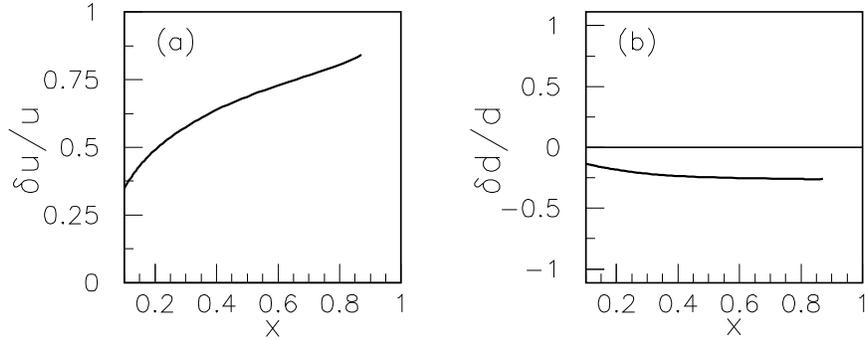}}
\end{center}
\caption[*]{\baselineskip 13pt The quark spin structure of the
transversely polarized proton at $Q^2=4~\rm{GeV}^2$: (a)~$\delta
u/u$ and (b)~$\delta d/d$. } \label{a03f4}
\end{figure}

The quark-diquark model with a simple form factor for the
nucleon-quark-diquark vertex  can provide good relations between
different quantities in which the uncertainties in the model can
be canceled between each other. It is impractical to expect a good
description of the absolute magnitude and shape for a physical
quantity. However, following Ref.~\cite{Efr00}, we may use some useful relations to connect the
unmeasured quantities with the measured quantities. For example,
we may use the following relation to connect the quark transverse
distributions with the quark unpolarized distributions
\begin{equation}
\begin{array}{clcr}
\delta u_{v}^N(x)
    =[u_v^N(x)-\frac{1}{2}d_v^N(x)]\hat{W}_S^{(u)}(x)
    -\frac{1}{6}d_v^N(x)\hat{W}_V^{(u)}(x);\\
\delta d_{v}^N(x)=-\frac{1}{3}d_v^N(x)\hat{W}_V^{(d)}(x),
\label{udv}
\end{array}
\end{equation}
with

\begin{equation}
\hat{W}_D^{(q)}(x)=\frac{\hat{A}_D^{(q)}(x)}{A_D^{(q)}(x)}.
\end{equation}
 We can use the unpolarized valence quark distributions
$u_v^N(x)$ and $d_v^N(x)$ at an initial scale  from one set of
quark distribution parametrizations as inputs to calculate the
quark transverse distributions with $\hat{W}_D^{(q)}(x)$ from the
diquark model calculation. In this way we can make more reliable
prediction for the absolute magnitude and shape of a physical
quantity than directly from the model calculation. It is worth to
emphasize that Eq.~(\ref{udv}) can only serve as the connection
between  the transverse polarized quark distributions and  the
unpolarized quark distributions at the initial scale ($\sim$
$1~\rm{GeV}^2$). The QCD evolution for the transverse quark
distributions is quite different from that for the unpolarized
quark distributions since the transverse distributions receive no
contribution from transversely polarized gluons. In the following
numerical calculations, the CTEQ5 set 1 parametrization
forms~\cite{CTEQ5} are adopted as  inputs for the unpolarized
quark distributions of the nucleon. In Fig.~\ref{a03f4}, we show
the  transversely polarized structure for the $u$ and $d$ quarks
in a transversely polarized proton at $Q^2=4 ~\rm{GeV}^2$.

\section{Azimuthal spin asymmetry in pion electroproduction}

Semi-inclusive pion production in  deep inelastic scattering
of leptons off a polarized nucleon target is a powerful tool for
providing information on  the spin structure of the nucleon
and on parton fragmentation. The HERMES
Collaboration~\cite{HERMES00} has recently reported
the measurement results of single-spin asymmetries
in the azimuthal distribution of
charged and neutral pions relative to the lepton scattering
 plane in deep-inelastic
scattering of positrons off longitudinally polarized protons.
It was found that the dependence of the single-spin azimuthal
asymmetry for $\pi^0$ production on the Bjorken scaling
variable $x$ is similar to the result for $\pi^+$ production,
while the $\pi^-$ asymmetry is consistent with zero within
present experimental uncertainties.

The analyzing power measured by HERMES  for unpolarized (U) beam and
longitudinally (L) polarized target is defined as
\begin{equation}
A_{UL}^W=\frac{\int \left[{\mathrm d} \phi\right] W(\phi)
\left\{N^+(\phi)-N^-(\phi)\right\}}{\frac{1}{2}\int \left[{\mathrm
d} \phi\right]\left\{ N^+(\phi)+N^-(\phi) \right\}}, \label{AP}
\end{equation}
where $W(\phi)=\sin \phi$ or $\sin 2 \phi$ is the weighting
function for picking up the Collins effect, and $N^+(\phi)$
($N^-(\phi)$) is the number of events for pion production as a
function of $\phi$ when the target is positively (negatively)
polarized. The analyzing powers for $\pi^+$, $\pi^0$ and $\pi^-$
have been measured~\cite{HERMES00}, and from the data there is
clear evidence for the non-zero values of $A_{UL}^{\sin \phi}$ for
$\pi^+$ and $\pi^0$, which indicates the azimuthal asymmetry.

\begin{figure}%[htb]
\begin{center}
\leavevmode {\epsfysize=3.5cm \epsffile{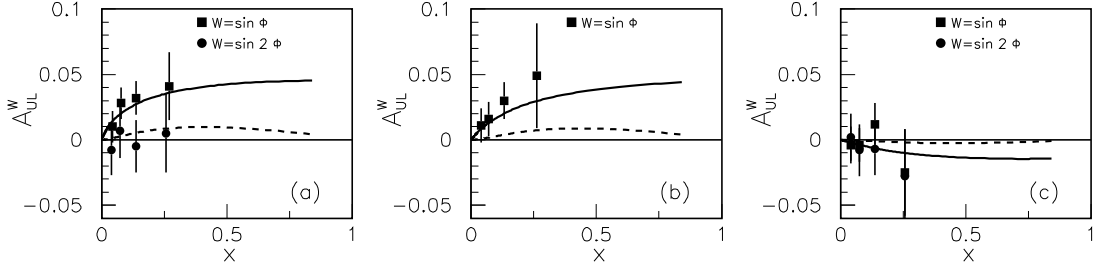}}
\end{center}
\caption[*]{\baselineskip 13pt The analyzing powers $A_{UL}^{\sin
\phi}$ and $A_{UL}^{\sin 2 \phi}$ for (a) $\pi^+$, (b) $\pi^0$ and
(c) $\pi^-$ of semi-inclusive pion production in deep inelastic
scattering of unpolarized positrons on the longitudinally
polarized proton target. The solid and dashed curves correspond to
$A_{UL}^{\sin \phi}$ and $A_{UL}^{\sin 2 \phi}$, respectively. The
input quark distributions are from CTEQ5 set 1 parametrization
forms~\cite{CTEQ5} at $Q^2=4$~GeV$^2$. }\label{a03f5}
\end{figure}

The analyzing power $A^{W}_{UL}$ is a function of the pion
fractional energy $z$, the Bjorken scaling variable $x$, and the
pion transverse momentum $P_\bot$. However, what we are interested
in the present investigation is the $x$ dependence of $A^{W}_{UL}$
since it can bring us information on the quark
transverse distributions of the nucleon. Based on a theoretical analysis
presented in Refs.~\cite{Mul96,Ams}, it was found~\cite{Efr00} 
that the analyzing powers $A_{UL}^{\sin \phi}$ 
and $A_{UL}^{\sin 2 \phi}$ are proportional to the ratios

\begin{equation}
\frac{\sum_q e^2_q \delta q^N (x) \left< \delta D^{\pi}_q(z)/z
\right>}{\sum_q e^2_q q^N(x) \left< D^{\pi}_q(z)\right>},
\label{AP1}
\end{equation}
and
\begin{equation}
\frac{ x ^2 \sum_q e^2_q \left( \int_x^1 {\mathrm d} \xi \delta
q^N (\xi) /\xi^2\right) \left< \delta D^{\pi}_q(z)\right>}{\sum_q
e^2_q q^N(x) \left< D^{\pi}_q(z)\right> }, \label{AP2}
\end{equation}
respectively. Here $e_q$ is the charge of the quark with flavor
$q$, $q^N(x)$ and $\delta q^N(x) $ are the quark unpolarized and
transverse distributions of the nucleon target, and $D^{\pi}_q(z)$
and $\delta D^{\pi}_q(z)$ are the fragmentation functions to the
pion $\pi$ from an unpolarized and transversely polarized quark
with flavor $q$. A further detailed theoretical analysis can be
found in Ref.~\cite{Ans00}. Therefore with the inputs of
$D^{\pi}_q(z)$ and $\delta D^{\pi}_q(z)$, we are able to get the
quark transverse distributions $\delta q^N(x)$ from the measured
analyzing powers. We will, following Ref.~\cite{Efr00}, consider
only the contributions from the favored fragmentation functions
$D^{\pi}_q$ and $\delta D^{\pi}_q$, {\it i.e.},
$D(z)=D^{\pi^{+}}_u(z)=D^{\pi^{+}}_{\bar{d}}(z)=2 D^{\pi^{0}}_u(z)
=2 D^{\pi^{0}}_{d}(z)=2 D^{\pi^{0}}_{\bar{u}}(z)
=2 D^{\pi^{0}}_{\bar{d}}(z)=
D^{\pi^{-}}_d(z)=D^{\pi^{-}}_{\bar{u}}(z)$ and similarly for
$\delta D^{\pi}_q$. The average values for $\left< D^{\pi}_q(z)
\right>$, $\left< \delta D^{\pi}_q(z) \right>$, $\left< \delta
D^{\pi}_q(z)/z \right>$, and the corresponding parameters are also
chosen the same as Ref.~\cite{Efr00}. Therefore, with the obtained
quark transverse distributions of the nucleon, we can  calculate
Eqs.~(\ref{AP1}) and (\ref{AP2}). The calculated results as shown
in Fig.~\ref{a03f5} are consistent with the HERMES experimental
data~\cite{HERMES00}. By contrasting the qualitative features of
the nucleon transversity distributions as shown in
Fig.~\ref{a03f4} with the results here, one can find that the
analyzing power $A_{UL}^{\sin \phi}$ for $\pi^+$ production 
is mainly controlled by $\delta u/u$. It is due to 
the charge factor for the $u$ quark and the fact 
that $\pi^+$ production is dominated by the $u \to \pi^+$ 
fragmentation process. In addition, the effect of the unfavored 
quark to pion fragmentation functions on 
$A_{UL}^{\sin \phi}$ for $\pi^+$ production is expected 
to be small. However, the situation of $A_{UL}^{\sin \phi}$ 
for $\pi^-$ production is a little complex since 
the contribution of unfavored quark to pion fragmentation 
functions might have significant effects on $\pi^-$ production. 
As for a detailed discussion, we would like to recommend our 
recent work in Ref.~\cite{MSY10}.
It is due to the fact that $\pi^+$ and $\pi^-$
productions are dominated by the $u \to \pi^+$ and $d \to \pi^-$
fragmentation processes, respectively. Therefore, we can extract
some useful information on the nucleon transversity distributions
by means of the analyzing powers $A_{UL}^{\sin \phi}$ for $\pi$
productions.

\section{Transverse $q \to \Lambda$ polarization transfers}

In addition to the azimuthal spin asymmetry in pion
electroproduction, the $\Lambda$ transverse polarization in the
current fragmentation region of charged lepton DIS on the
transversely polarized nucleon target can also provide information
on the quark transverse distribution of the target. For $\Lambda$
production in the current fragmentation region along the virtual
photon direction~\cite{MSSY5}, the spin transfer to the
transversely polarized $\Lambda$ can be  written
as~\cite{Jaf96,Art90}
\begin{equation}
\hat{A}^{\Lambda}(x,z)= \frac{\sum\limits_{q} e_q^2 \delta
q^N(x,Q^2) \delta D_q^\Lambda(z,Q^2) } {\sum\limits_{q} e_q^2 q^N
(x,Q^2) D^\Lambda_q(z,Q^2)}~ \label{TST}
\end{equation}
for charged lepton DIS on a transversely polarized nucleon $N$
target. A detailed discussion about the transverse spin transfer
has been given in Ref.~\cite{MSSY6}. Now we can predict this 
transverse spin transfer with
the obtained quark distributions $q^N(x)$, $\delta q^N(x)$ for the
nucleon and  the  $q \to \Lambda$ fragmentation functions
$D_q^{\Lambda}(z)$ and $\delta D_q^{\Lambda}(z)$.

\begin{figure}%[htb]
\begin{center}
\leavevmode {\epsfysize=5.6cm \epsffile{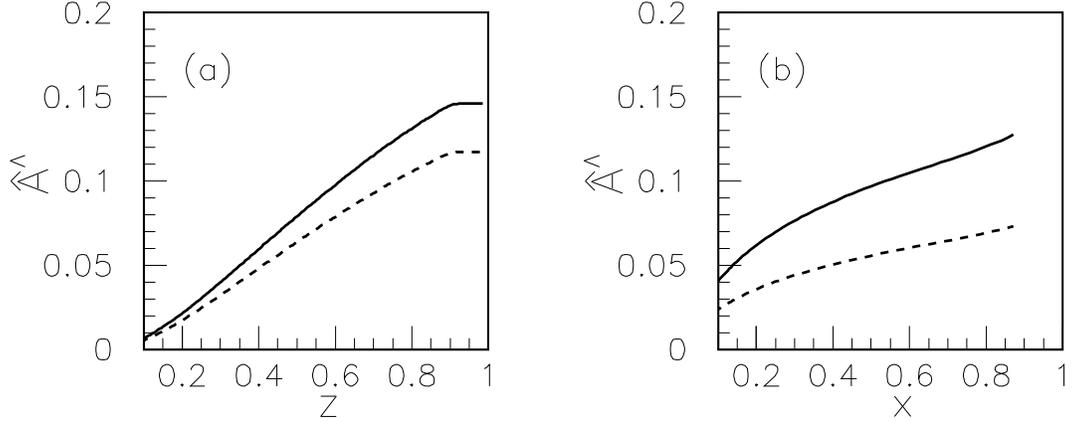}}
\end{center}
\caption[*]{\baselineskip 13pt The $x$-integrated and
$z$-integrated spin transfers $\hat{A}^{\Lambda}(x,z)$ of
$\Lambda$ production in charged lepton DIS process on the
transversely polarized proton  target, with the integrated $x$ or
$z$ range of $0.6 \to 1$ for the solid curves and $0.3 \to 1$ for
the dashed curves at $Q^2= 4~\rm{GeV}^2$. }\label{a03f6}
\end{figure}

The $x$-integrated and $z$-integrated spin transfers
$\hat{A}^{\Lambda}(x,z)$   are presented in Fig.~\ref{a03f6}. The
results can be easily understood by  contrasting  the qualitative
features of the transversely polarized $q \to \Lambda$
fragmentation functions as shown in Fig.~\ref{a03f2}(b) with the
$z$ dependence  of $\hat{A}^{\Lambda}$ in Fig.~\ref{a03f6}(a) and the
features of the nucleon transversity  distributions in
Fig.~\ref{a03f4}(a) with the $x$  dependence of  $\hat{A}^{\Lambda}$ in
Fig.~\ref{a03f6}(b). We will have some more detailed discussions
in the next section. After all, it turns out that the
$x$-integrated and $z$-integrated $\hat{A}^{\Lambda}(x,z)$ can
provide us more information on the $q \to \Lambda$ fragmentation
functions and the quark transverse distributions of the nucleon,
respectively.

\section{Discussion and summary}

The $\Lambda$ polarization $P^{\Lambda}$ in $e^+e^-$
annihilation at the Z-pole provides  us information on
the polarized $s$ quark to
$\Lambda$ fragmentation function since it is dominated by  the $s
\to \Lambda$ fragmentation. The spin transfer $A^{\Lambda}$ in $\Lambda$
electroproduction is dominated by the $u \to \Lambda$
fragmentation due to the charge factor for the $u$ quark, {\it
i.e.,} Eq.~(\ref{DL}) can be approximated by

\begin{equation}
A^{\Lambda} \simeq \frac{ \Delta D_u^\Lambda(z,Q^2)}
{D^\Lambda_u(z,Q^2)}.~ \label{DLp}
\end{equation}
The $d$ quark spin transfer to the $\Lambda$ is expected to be
equal to that of the $u$ quark because of the isospin symmetry.
Consequently, $\Lambda$ electroproduction in the current
fragmentation region is most sensitive to the ratio $\Delta
D_u^\Lambda/D^\Lambda_u=\Delta D_d^\Lambda/D^\Lambda_d$.
Therefore, our consistent descriptions of the experimental data on
the $\Lambda$ polarization $P^{\Lambda}$ and the spin transfer
$A^{\Lambda}$ indicate that the $q \to \Lambda$ fragmentation
functions obtained in the diquark model are reasonable. According
to Eq.~(\ref{DLp}), the available HERMES experimental
data~\cite{HERMES} on the spin transfer suggests that the  $u$ and
$d$ quark to the $\Lambda$ fragmentation functions  are likely
positive polarized in medium and large $z$ region. Furthermore,
the very recent NOMAD data~\cite{NOMAD} on the $\Lambda$
polarization in the neutrino DIS process, which is also dominated
by the $u \to \Lambda$ fragmentation, supports the above
information. However, despite the experimental uncertainties, the
E665 experiment~\cite{E665} seems to show negative $\Lambda$
polarizations in low $z$ region where the sea quark contribution
dominates. Unfortunately, the polarized sea quark to $\Lambda$
fragmentation can neither be constructed within the framework of
the diquark model nor be extracted from the available experimental
data. A further study on the small $z$ region behavior of the $q
\to \Lambda$ polarization transfer is required.

 The analyzing power $A_{UL}^{\sin \phi}$ for  $\pi^+$ production
is dominated by the $u$ quark from the fragmentation of $u \to
\pi^+$. Therefore, the analyzing power  $A_{UL}^{\sin \phi}$ for
$\pi^+$ production  with summed over $z$ can provide information
on the $u$ quark transverse distributions of the target nucleon.
The increase of $A^{\sin \phi}_{UL}$ with increasing $x$ as shown
in the HERMES data~\cite{HERMES00} for $\pi^+$ and $\pi^0$
productions  suggests that single-spin asymmetries are with
valence quark distributions. On the other hand, the spin transfer
 $\hat{A}^{\Lambda}(x,z)$ of $\Lambda$ production in charged lepton
 DIS process on the transversely polarized proton  target is also
 dominated by the $u \to \Lambda$ spin transfer
 due to the charge factor for the $u$ quark.  Eq.~(\ref{TST}) can
 be  approximately reduced as

 \begin{equation}
\hat{A}^{\Lambda}(x,z)\simeq \frac{ \delta u^N(x,Q^2) \delta
D_u^\Lambda(z,Q^2) } {u^N (x,Q^2) D^\Lambda_u(z,Q^2)}.~
\label{TSTp}
\end{equation}
Hence, with $\delta u^N(x)/u^N (x)$ from the power analyzing
$A_{UL}^{\sin \phi}$ for  $\pi^+$ production, $\delta
D_u^\Lambda/D^\Lambda_u$ can be extracted from  the $x$-integrated
$\hat{A}^{\Lambda}(x,z)$. We also present the spin transfers
integrated over different $z$ regions in Fig.~\ref{a03f6}. We find
that the $x$ dependence of the $z$-integrated
$\hat{A}^{\Lambda}(x,z)$ is mainly shaped  by  $\delta
u^N(x)/u^N(x)$ (see Fig.~\ref{a03f4}(a) and Fig.~\ref{a03f6}(b)),
as well as the $x$ dependence of the analyzing power $A_{UL}^{\sin
\phi}$ for the  $\pi^+$ production (see Fig.~\ref{a03f4}(a) and
Fig.~\ref{a03f5}(a)). The magnitude values of the $z$-integrated
$\hat{A}^{\Lambda}(x,z)$ and the analyzing power $A_{UL}^{\sin
\phi}$ for the $\pi^+$ production are sensitive to the
$z$-integrated value of the fragmentation functions for the
produced $\Lambda$ and $\pi^+$, respectively. Therefore, in order
to make a mutual check for the transversely polarized $q \to
\Lambda$ fragmentation functions and the transverse quark
distributions of the nucleon, it is very significant to perform
the measurement of the transverse spin transfer
$\hat{A}^{\Lambda}(x,z)$ over various ranges of $x$ and $z$, as
well as further precision measurement of $A_{UL}^{\sin \phi}$ for
pion production.

In summary, based on the available unpolarized $\Lambda$
production data in $e^+e^-$ annihilation, we extracted  a set of
$q \to \Lambda$ fragmentation functions with the flavor and spin
structure from the quark-diquark model.  It is found that the
longitudinal spin structure of the obtained $\Lambda$
fragmentation functions is supported by the all  available
experimental data on the longitudinal $\Lambda$ polarization.
Within the same framework, we proposed a set of transverse quark
distributions of the nucleon and obtained consistent descriptions
of the available HERMES data on the azimuthal spin asymmetry in
pion electroproduction. In order to make a mutual check of the
quark  transverse distributions of the nucleon and the
transversely polarized  quark to $\Lambda$ fragmentation
functions, we emphasized the significance of the measurement on
spin transfer to the transversely polarized $\Lambda$ in the
charged lepton DIS on a transversely polarized nucleon target.
The investigation on the $q \to \Lambda$ fragmentation
functions and the quark transverse distributions of the
nucleon is important for enriching our knowledge of hadron
structure and hadronization mechanism.

{\bf Acknowledgments: } I am grateful to P. J. Mulders and J.
Rodrigues for their kindness correspondence about the
quark-diquark model. This work is stimulated by my other
cooperative work with Bo-Qiang Ma, Ivan Schmidt, and Jacques
Soffer. I also would like to express my great thanks to them for
their encouragements and valuable comments.  In
addition, this work is partially supported by National Natural
Science Foundation of China under Grant Number 19875024 and by
Fondecyt (Chile) project 3990048.

\end{document}